\documentclass[aps,prd,preprint,superscriptaddress,tightenlines,nofootinbib]{revtex4}

\usepackage{graphicx}
\usepackage{dcolumn}
\usepackage{bm}

\usepackage{amssymb,amsmath}
\usepackage{amsthm}
\usepackage{amsfonts}
\usepackage{avant}

\begin{document}

\preprint{CLNS 07/2012}       
\preprint{CLEO 07-16}         

\title{\boldmath{Absolute Branching Fractions of Cabibbo-Suppressed $D \to K\bar{K}$ Decays}}

\author{G.~Bonvicini}
\author{D.~Cinabro}
\author{M.~Dubrovin}
\author{A.~Lincoln}
\affiliation{Wayne State University, Detroit, Michigan 48202, USA}
\author{J.~Rademacker}
\affiliation{University of Bristol, Bristol BS8 1TL, UK}
\author{D.~M.~Asner}
\author{K.~W.~Edwards}
\author{P.~Naik}
\author{J.~Reed}
\affiliation{Carleton University, Ottawa, Ontario, Canada K1S 5B6}
\author{R.~A.~Briere}
\author{T.~Ferguson}
\author{G.~Tatishvili}
\author{H.~Vogel}
\author{M.~E.~Watkins}
\affiliation{Carnegie Mellon University, Pittsburgh, Pennsylvania 15213, USA}
% ...
\collaboration{CLEO Collaboration} 
\noaffiliation

\date{June 2, 2008}

\begin{abstract} 
Using 281 $\rm{pb}^{-1}$ of data collected with the CLEO-c detector at the $\psi (3770)$ resonance, we have studied Cabibbo-suppressed decays of $D$ mesons to final states with two kaons.  We present results for the absolute branching fractions of the modes $D^{0} \to K^{+}K^{-}$, $D^{0} \to K^{0}_{S}K^{0}_{S}$, and $D^{+} \to K^{+}K^{0}_{S}$.  We measure ${\cal B}(D^{0} \to K^{+}K^{-})=(4.08 \pm 0.08 \pm 0.09) \times 10^{-3}$, ${\cal B}(D^{0} \to K^{0}_{S}K^{0}_{S}) =(1.46 \pm 0.32 \pm 0.09) \times 10^{-4}$, and ${\cal B}(D^{+} \to K^{+}K^{0}_{S})=(3.14 \pm 0.09 \pm 0.08) \times 10^{-3}$.  We also determine the ratio ${\cal B}(D^{0} \to K^{+}K^{-})/{\cal B}(D^{0} \to \pi^{+}\pi^{-}) = 2.89 \pm 0.05 \pm 0.06$.  For each measurement, the first uncertainty is statistical and the second uncertainty is systematic.
\end{abstract}

\pacs{13.25.Ft}
\maketitle

Detailed studies of the decay of charmed mesons provide an opportunity to probe the interplay between the strong and weak force.  One such example is the study of the contributions of $W$-exchange diagrams and final state interactions~\cite{CLEO962}.  In particular, the branching ratio ${\cal B}(D^{0} \to K^{+}K^{-})/{\cal B}(D^{0} \to \pi^{+}\pi^{-})$ exposes large SU(3) flavor symmetry breaking effects.  To first order one would expect a branching ratio close to unity, but the most recent studies show the ratio to be $2.760 \pm 0.040 \pm 0.034$~\cite{cdf05c}.  The direct weak decay $D^{0} \to K^{0}\bar{K}^{0}$ is expected to occur primarily via two $W$-exchange diagrams, which in a four-quark model under SU(3) is expected to be zero due to destructive interference~\cite{kskstheory2}.  Hence, substantial contributions to the rate for $D^{0} \to K^{0}\bar{K}^{0}$ are expected from rescattering in such channels as $D^0 \to K^{*+}K^{*-}$, $D^0 \to \pi^+ \pi^-$, and $D^0 \to \rho^+ \rho^-$~\cite{kskstheory, eeg, dai}.  Theoretical estimates of SU(3)-breaking effects give branching fractions ranging from a few parts in $10^4$~\cite{eeg} up to $0.3\%$~\cite{kskstheory}.  Here we study the decay mode $D^0 \to K^0_S K^0_S$, whose branching fraction should be half that of $D^0 \to K^0 \bar{K}^0$.  Note that the decay mode $D^0 \to K^0_S K^0_L$ is forbidden by $CP$ conservation.  Understanding SU(3) flavor symmetry breaking is important in attempting to quantify Standard Model contributions to $D^0-\bar{D}^0$ mixing~\cite{ksktheory}.  The decay mode $D^{+} \to K^{+}K^{0}_{S}$ is useful for the estimation of SU(3)-violating effects in the $D$ meson system~\cite{CLEO0310}.  Thus, a precise measurement of the absolute branching fraction of $D \to K\bar{K}$ modes will test our current understanding of the Standard Model.

We use a 281 $\rm{pb}^{-1}$ data sample of $e^{+}e^{-}$ collisions at a center-of-mass energy of approximately 3770~MeV produced at the Cornell Electron Storage Ring and recorded with the CLEO-c detector.  The CLEO-c detector is a general purpose solenoidal detector that includes a tracking system for measuring momenta and specific ionization ($dE/dx$) of charged particles, a ring imaging Cherenkov detector for particle identification, and a CsI calorimeter for detection of photons.  The details of the CLEO-c detector have been described elsewhere~\cite{cleoc1, cleoc2, cleoc3}.

We are interested in events of the type $e^{+}e^{-} \to \psi (3770) \to D\bar{D}$.  The $D$ mesons have energy equal to the beam energy since the $\psi (3770)$ resonance is below the kinematic threshold for $D\bar{D}\pi$ production.  After selecting candidate particles that may be the daughters of a $D$ meson decay, we define the following two variables, $\Delta E \equiv \sum_i E_{i} - E_{\rm{beam}}$ and $M_{\rm BC} \equiv \sqrt{E^2_{\rm{beam}} - | \sum_i \mbox{\textbf{p}}_{i} |^{2}}$, where $E_{i}$ and \( \mbox{\textbf{p}}_{i} \) are the energies and momenta respectively of the $D$ meson daughters and $E_{\rm{beam}}$ is the beam energy.  If the event is reconstructed correctly, $\Delta E$ should be consistent with zero and the beam constrained mass $M_{\rm BC}$ should be consistent with the mass of the $D$ meson.

In selecting events, charged tracks are required to satisfy criteria based on the track quality~\cite{CBX0506}.  They must also be consistent with coming from the interaction point in three dimensions.  Pions and kaons are identified by consistency with the expected $dE/dx$ and, when available, ring imaging Cherenkov information.  We also require that $\left| \Delta E \right|$ be less than 20~MeV.  Other selection criteria are mode dependent as described below.  Charge-conjugate modes are implied throughout.

In selecting events for the mode $D^{0} \to K^{+}K^{-}$, we veto events that contain leptons, e.g. cosmic rays and Bhabha events, and furthermore require $\cos \theta > -0.9$ for $K^-$ and $\cos \theta < 0.9$ for $K^+$, where $\theta$ is the angle between the $e^+$ direction and the kaon momentum direction.  The $\cos \theta$ criterion is useful to further suppress radiative Bhabha background events that have the same energy loss at about 800~MeV as the kaons in $D^{0} \to K^{+}K^{-}$.

In selecting events for the mode $D^{+} \to K^{+}K^{0}_{S}$, we require the $K^{0}_{S}$ candidate invariant mass to be within $12{~\rm{MeV}}/c^2$, or roughly $4.5 \sigma$, of the known $K^{0}_{S}$ invariant mass~\cite{pdg} and require the $K^0_S$ candidate to have a flight distance significance $(L/\sigma_L)$ greater than two in order to separate true $K^0_S$ particles from pairs of pions produced at the interaction point.

\begin{figure}
\begin{center}
 \includegraphics*[width=0.45\textwidth]{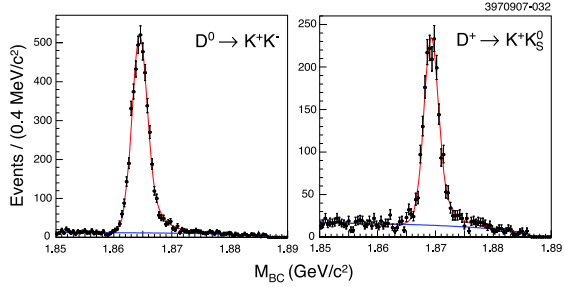}
 \end{center}
\caption{$M_{\rm{BC}}$ plots for data, left to right: $D^{0} \to K^{+} K^{-}$ and $D^{+} \to K^{+} K^{0}_{S}$.}
\label{fig:data1}
\end{figure}

The mode $D^{0} \to K^{0}_{S}K^{0}_{S}$ has significant backgrounds, mostly from the mode $D^{0} \to K^{0}_{S} \pi^{+} \pi^{-}$ where the $\pi^+$ and the $\pi^-$ have an invariant mass consistent with the $K^{0}_{S}$ mass.  Therefore, similar to our analysis of the mode $D^{+} \to K^{+}K^{0}_{S}$, we require the flight distance significance for each of the $K^{0}_{S}$ candidates to be greater than two.  Furthermore, we perform a background subtraction where we take the region $|M_{\pi^{+}\pi^{-}}-M_{K^{0}_{S}}|<7.5$~MeV$/c^2$ to be our signal region and take the region 20~$>|M_{\pi^{+}\pi^{-}}-M_{K^{0}_{S}}|>9$~MeV$/c^2$ to be our background region, where $M_{\pi^{+}\pi^{-}}$ is the reconstructed $K^{0}_{S}$ candidate invariant mass and $M_{K^{0}_{S}}$ is the known $K^{0}_{S}$ invariant mass~\cite{pdg}.  This will be described in detail later.

\begin{figure*}
\begin{center}
 \includegraphics*[width=0.85\textwidth]{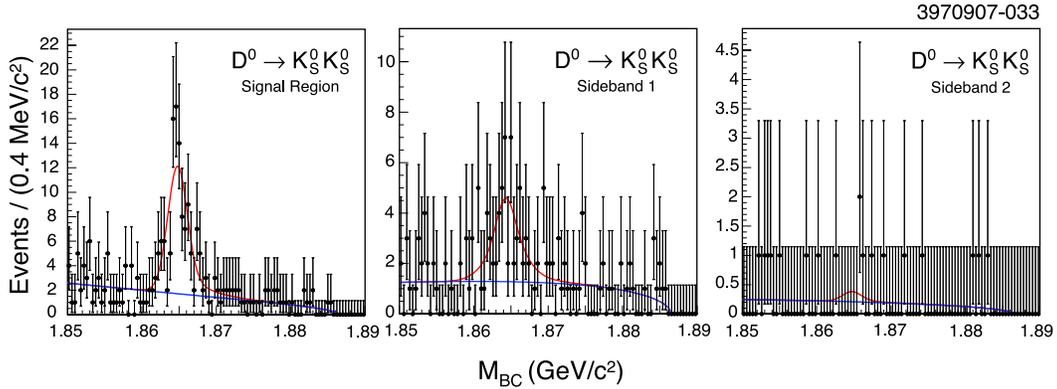}
 \end{center}
\caption{$D^{0} \to K^{0}_{S}K^{0}_{S}$ plots for data, left to right: $M_{\rm{BC}}$ fits for the signal region for $D^{0} \to K^{0}_{S}K^{0}_{S}$, sideband region 1 for $D^{0} \to K^{0}_{S}K^{0}_{S}$ and sideband region 2 for $D^{0} \to K^{0}_{S}K^{0}_{S}$.  The sideband regions are shown in Fig.~\ref{fig:data3}.}
\label{fig:data2}
\end{figure*}

The branching fractions are calculated using ${\cal B}={Y / 2\epsilon N_{D\bar D}}$, where $Y$ is the measured signal yield, $\epsilon$ is the efficiency obtained from Monte Carlo simulations, and $N_{D \bar D}$ is the produced number of $D \bar D$ pairs (charged or neutral depending on the decay mode) in our data sample.  We use $N_{D \bar D}$ from Ref.~\cite{CBX0506}.

To determine the measured yields, we perform an unbinned maximum likelihood fit to a signal shape and one background component to extract signal yields from the $M_{\rm{BC}}$ distributions.  The signal shape includes the effects of beam energy spread, initial state radiation, the line shape of the $\psi(3770)$, and reconstruction resolution~\cite{CBX0506}.  The background is described by an ARGUS function~\cite{argus}, which models combinatorial contributions.  We have validated our analysis procedure by studying simulated $D\bar{D}$ events in a sample 40 times the size of our data set and we reproduce the input branching fractions to within statistical uncertainties.  Our fits to the data are shown in Figs.~\ref{fig:data1} and~\ref{fig:data2}, and our efficiencies, along with the fitted yields, are given in Table~\ref{Ta:data}.

\begin{table}
    \caption{Efficiencies and yields with their statistical uncertainties.  The yield for $D^{0} \to K^{0}_{S}K^{0}_{S}$ is the yield after background subtraction.}
\begin{center}
    \begin{tabular}{ccc} \hline \hline
    Mode & Efficiency ($\%$) & Yield \\ \hline
    $D^{0} \to K^{+}K^{-}$ & $56.47 \pm 0.33$ & $4746 \pm 74$ \\
    $D^{0} \to K^{0}_{S}K^{0}_{S}$ & $22.75 \pm 0.22$ & $68 \pm 15$ \\
    $D^{+} \to K^{+}K^{0}_{S}$ & $38.29 \pm 0.28$ & $1971 \pm 51$ \\ \hline \hline
    \end{tabular}
\label{Ta:data}
\end{center}
\end{table}

\begin{figure}
\begin{center}
 \includegraphics*[width=0.45\textwidth]{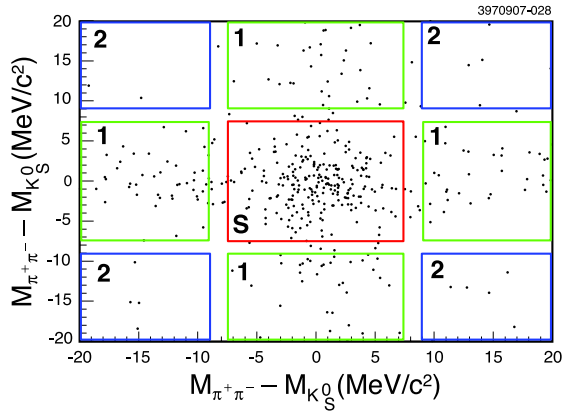}
 \end{center}
\caption{$(M_{\pi^+ \pi^-}-M_{K^0_S})$ vs $(M_{\pi^+ \pi^-}-M_{K^0_S})$ for $D^{0} \to K^{0}_{S}K^{0}_{S}$ candidates.  The region labeled ``S'' is the signal region, the regions labeled ``1'' are the sideband 1 region, and the regions labeled ``2'' are the sideband 2 region.}
\label{fig:data3}
\end{figure}

The scatter plot for the two $K^{0}_{S}$ candidate invariant masses in data in the region $|M_{\pi^{+}\pi^{-}}-M_{K^{0}_{S}}|<20$~MeV$/c^2$ is shown in Fig.~\ref{fig:data3}, where $M_{\pi^{+}\pi^{-}}$ is the reconstructed $K^{0}_{S}$ candidate invariant mass and $M_{K^{0}_{S}}$ is the known $K^{0}_{S}$ invariant mass~\cite{pdg}.  The order of the two $K^{0}_{S}$ candidates is essentially random as it is determined by the order the tracks were found in the event.  We expect the bulk of the signal to be in the central (signal) region which we define to be $|M_{\pi^{+}\pi^{-}}-M_{K^{0}_{S}}|<7.5$~MeV$/c^2$ for the two $K^{0}_{S}$ candidates.  In order to account for $D^{0} \to K^{0}_{S} \pi^{+} \pi^{-}$ and other background events entering into the signal region of $D^{0} \to K^{0}_{S}K^{0}_{S}$, we perform a background subtraction.
We then take the four sideband regions where one $K^{0}_{S}$ candidate has 20~$>|M_{\pi^{+}\pi^{-}}-M_{K^{0}_{S}}|>9$~MeV$/c^2$ and the other $K^{0}_{S}$ candidate has $|M_{\pi^{+}\pi^{-}}-M_{K^{0}_{S}}|<7.5$~MeV$/c^2$.  We call this region sideband 1.  Finally we take the four corner regions where both $K^{0}_{S}$ candidates satisfy 20~$>|M_{\pi^{+}\pi^{-}}-M_{K^{0}_{S}}|>9$~MeV$/c^2$.  We call this region sideband 2.  The number of candidate events in each region is determined by a fit to the $M_{\rm BC}$ distribution of the $D$ meson candidates and are found to be $S = 95 \pm 13$, $S_1 = 40 \pm 11$, and $S_2 = 2 \pm 3$ for the signal region, sideband 1, and sideband 2 respectively.  We let $r_1$ denote the ratio of the area in the signal region to the sideband region 1 and $r_2$ denote the ratio of the area in the signal region to the sideband region 2.  Then the total yield $Y$ is given by $Y=S-2r_{1}S_{1}+r_{2}S_{2}$.  Note that $S_1$ is the measured yield in two bands, so after rescaling $S_1$ with $r_1$, a factor of two is needed in the second term on the right hand side to account for the fact that two bands are overlapping in the signal region.  The third term is added since the second term oversubtracts the uniform background, but we find that $S_2$ is much smaller than $S$ and $S_1$.  We have $r_1=225/660$ and $r_2=225/484$.

\begin{table*}
    \caption{Measured branching fractions and Particle Data Group (PDG) values~\cite{pdg}; the first uncertainty is statistical and the second uncertainty is systematic.}
\begin{center}
    \begin{tabular}{ccc} \hline \hline
    Mode & Our Result & PDG 2007 \\ \hline
    ${\cal B}(D^{0} \to K^{+}K^{-})$ & $(4.08 \pm 0.08 \pm 0.09) \times 10^{-3}$ & $(3.85 \pm 0.09) \times 10^{-3}$ \\
    ${\cal B}(D^{0} \to K^{0}_{S}K^{0}_{S})$ & $(1.46 \pm 0.32 \pm 0.09) \times 10^{-4}$ & $(3.6 \pm 0.7) \times 10^{-4}$ \\
    ${\cal B}(D^{+} \to K^{+}K^{0}_{S})$ & $(3.14 \pm 0.09 \pm 0.08) \times 10^{-3}$ & $(2.95 \pm 0.19) \times 10^{-3}$ \\ \hline \hline
    \end{tabular}
\label{Ta:bf}
\end{center}
\end{table*}

We have studied various systematic uncertainties in our analysis.  These include particle identification efficiency, tracking efficiency, uncertainties in the fitting procedure, and the uncertainties due to the $\Delta E$ selection criterion.  Using techniques outlined in Ref.~\cite{CBX0506}, the systematic error in the absolute tracking efficiency for pions (kaons) was determined to be $0.3 \%$ ($0.7 \%$).  We have also determined that our Monte Carlo simulation accurately models the identification efficiency of pions, kaons, and $K^0_S$ to within $0.3 \%$, $0.2 \%$, and $1.9\%$ respectively.  To determine the systematic errors due to the fitting procedure, we varied the background shape and the momentum resolution in the fit and compared the variations in the fit yield.  These have been determined to be $0.8\%$, $0.6\%$, and $4.1\%$ for each of the modes $D^{0} \to K^{+}K^{-}$, $D^{+} \to K^{+}K^{0}_{S}$, and $D^{0} \to K^{0}_{S}K^{0}_{S}$ respectively.  To determine the systematic error due to the $\Delta E$ criterion, we varied the $\Delta E$ criterion and studied its effect on the yield.  This has a negligible effect for the modes $D^{0} \to K^{+}K^{-}$, $D^{+} \to K^{+}K^{0}_{S}$, and for the mode $D^{0} \to K^{0}_{S}K^{0}_{S}$ we assign a systematic error of $2.0\%$.  Finally, using a detailed Monte Carlo simulation of the mode $D^{0} \to K^0_S \pi^+ \pi^-$, we estimated the effect of non-linearity in the background shape for the background subtraction in the mode $D^{0} \to K^{0}_{S}K^{0}_{S}$ and found it to be negligible.

The measured branching fractions are summarized in Table~\ref{Ta:bf}.  We use the $D\bar{D}$ yields $N_{D^{0}\bar{D^{0}}}=(1.031 \pm 0.008 \pm 0.013) \times 10^{6}$ and $N_{D^{+}D^{-}}=(0.819 \pm 0.008 \pm 0.010) \times 10^{6}$ as measured in Ref.~\cite{CBX0506} to determine our branching fractions.  We also determined the ratio of branching fractions ${\cal B}(D^{0} \to K^{+}K^{-})$ to ${\cal B}(D^{0} \to \pi^{+}\pi^{-})$, where the latter branching fraction has previously been measured using the same data sample~\cite{pipi}, to be ${\cal B}(D^{0} \to K^{+}K^{-})/{\cal B}(D^{0} \to \pi^{+}\pi^{-}) = 2.89 \pm 0.05 \pm 0.06$, consistent with the previous measured value of ${\cal B}(D^{0} \to K^{+}K^{-})/{\cal B}(D^{0} \to \pi^{+}\pi^{-}) = 2.760 \pm 0.040 \pm 0.034$~\cite{cdf05c}.  The modes $D^{0} \to K^{+}K^{-}$ and $D^{+} \to K^{+}K^{0}_{S}$ are consistent with the PDG average.  Finally, for the $D^{0} \to K^{0}_{S}K^{0}_{S}$ mode, our errors are significantly smaller than previous measurements, while our branching fraction is about a factor of 2.4 (2.7 standard deviations) smaller~\cite{pdg}.

We gratefully acknowledge the effort of the CESR staff
in providing us with excellent luminosity and running conditions.
D.~Cronin-Hennessy and A.~Ryd thank the A.P.~Sloan Foundation.
This work was supported by the National Science Foundation,
the U.S. Department of Energy,
the Natural Sciences and Engineering Research Council of Canada, and
the U.K. Science and Technology Facilities Council.

\end{document}